\documentclass[aps,prl,reprint,preprintnumbers]{revtex4-1}
\usepackage{graphicx}
\usepackage{bm}
\usepackage{epsfig}
\usepackage{ulem}
\usepackage{color}
\usepackage{mathrsfs}
\usepackage{dcolumn}
\usepackage{setspace}
\usepackage{array}
\usepackage{amsmath}
\usepackage{amssymb}
\usepackage{dsfont}
\usepackage{gensymb}
\usepackage{dcolumn}
\usepackage{multirow}
\usepackage{bibentry,natbib}
\usepackage{booktabs}
\begin{document}
\bibliographystyle{unsrt}
\title{Chiral Casimir Forces: Repulsive, Enhanced, Tunable}

\author{Qing-Dong Jiang$^1$, Frank Wilczek${^1}{^2}{^3}{^4}$}
\affiliation{{}\\ $^1$Department of Physics, Stockholm University, Stockholm SE-106 91 Sweden\\
$^2$Center for Theoretical Physics, Massachusetts Institute of Technology, Cambridge, Massachusetts 02139 USA\\
$^3$Wilczek Quantum Center, Department of Physics and Astronomy, Shanghai Jiao Tong University, Shanghai 200240, China\\
$^4$Department of Physics and Origins Project, Arizona State University, Tempe AZ 25287 USA}
\begin{abstract}
Both theoretical interest and practical significance attach to the sign and strength of Casimir forces.  A famous, discouraging no-go theorem states that ``The Casimir force between two bodies with reflection symmetry is always attractive." Here we identify a loophole in the reasoning, and propose a universal way to realize repulsive Casimir forces.  We show that the sign and strength of Casimir forces can be adjusted by inserting optically active or gyrotropic media between bodies, and modulated by external fields.
\end{abstract}
\preprint{MIT-CTP/5002}
\maketitle


\textit{Introduction}: The Casimir effect is one of the best known macroscopic manifestations of quantum field theory, and has attracted interest since its first discovery \cite{Casimir}.  
The original version of Casimir effect is an attractive force between two ideal, uncharged metal plates in vacuum.  Later on, Lifshitz \textit{et al.}, derived a general formula for the Casimir force between between materials described by dielectric response functions in this geometry \cite{Lifshitz}.  In their formula, the Casimir force between material 1 and material 2 across medium 3 is proportional to a summation of terms with differences in material dielectric functions
\begin{eqnarray}\label{eq1}
-\bigl(\epsilon_1(\omega)-\epsilon_3(\omega)\bigr)\bigl(\epsilon_2(\omega)-\epsilon_3(\omega)\bigr)
\end{eqnarray}
over frequency $\omega$, where $\epsilon_i$ is the dielectric function for material $i$ ($i=1,2,3$). Between two  like materials,  $\epsilon_1=\epsilon_2$, these terms are always negative and correspond to attractive Casimir force, regardless the mediating material 3.  A famous generalization of this result states that objects made of the same isotropic material always attract for reflection symmetric geometries (but arbitrary shapes) in  vacuum \cite{oken}, or for a wide class of intermediate materials, as we will review presently.  This strong theorem appears to rule out many convenient possibilities for realizing repulsive Casimir forces.

Yet in principle the Casimir force can be repulsive. In recent years, people have devoted substantial efforts to realizing repulsive Casimir forces, especially with a view toward applications to nano-devices and colloids, which can contain nearby parts  that one wants to keep separate. In fact, repulsive Casimir forces have been proposed in several special cases \cite{Boyer,Levin,Leonhardt}, and have even been observed experimentally \cite{jnmunday}. In this experiment, the authors measured the Casimir force between gold (solid) and silica (solid) mediated by bromobenzene (liquid), of which the dielectric functions satisfy $\epsilon_1>\epsilon_3>\epsilon_2$.   In recent years, the repulsive Casimir force has been also proposed in various topological and metamaterials \cite{Tse2012,rosa}. However, all these proposals give tiny repulsive Casimir forces (compared to the Casimir force between metals), and demand particular parameters of materials, or particular shapes of materials, making experimental realization challenging and somewhat awkward.

In this paper, we do two things.  First, we identify an important loophole in the central no-go theorem \cite{oken} on Casimir forces.  It arises when there is an intervening ``lubricant'' material with no symmetry between left- and right-circular polarized photons (i.e., a chiral material).  Optically active materials, which break spatial parity but preserve time reversal, are not rare, and provide good candidates. Second, we explicitly calculate the Casimir force between similar objects separated by a chiral medium (see figure 1).    We find that the Casimir force can, as a function of distance, oscillate between attractive and repulsive, and that it can be tuned by application of an external magnetic field.

\begin{figure}[!htb]
\centering
\includegraphics[height=3.5cm, width=6cm, angle=0]{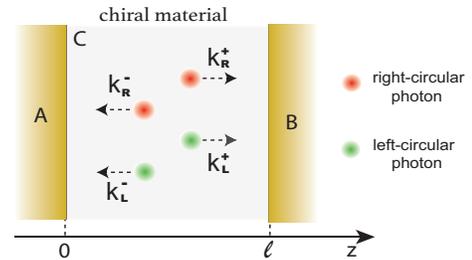}
\caption{Schematic illustration of chiral Casimir effect. Two parallel, uncharged plates (A \& B) are placed at a distance $l$ separated by chiral material C.  The red dots and green dots represent right-circular polarized photons and left-circular polarized photons.  The arrows indicate the propagating directions of chiral photons. $\rm k_{R(L)}^{\pm}$ represent velocity of chiral photons, where superscript $\pm$ correspond to their propagating directions, and the subscript $\rm R/L$ correspond to their chirality. }\label{figure1}
\end{figure}


\textit{Identifying the loophole}: To begin, we briefly review the ``Casimir'' energy in massless free scalar field theory.  The action of free scalar field in a dielectric medium is 
\begin{eqnarray}\label{eq2}
S&=&\int dx \mathcal L \nonumber \\
&=&\frac{1}{2} \int d\bold x \frac{d\omega}{(2\pi)} \phi^*(\bold r,\omega)\left( \chi \omega^2+\nabla ^2\right)\phi(\bold r,\omega)
\end{eqnarray}
where $\phi(\bold r,\omega)=\int_{-\infty}^{\infty} dt \,\phi(\bold r, t)e^{i\omega t}$ is the Fourier conjugate of $\phi(\bold r, t)$, and $\chi$ is the dielectric function of the corresponding system. (As a default, we adopt units with $\hbar = c = 1$. Note that $\chi=1$ in vacuum and $\chi\neq 1$ in materials.) 
Assume that two dielectric bodies A, B are separated by a medium C.  We write the space-dependent dielectric function as $\chi(\bold r,\omega)=\chi_0(\omega)+\chi^{\prime}(\bold r,\omega)$ with $\chi^{\prime}(\bold r,\omega)=0$ in region C but $\chi^{\prime}_{\rm A \, or \, B}(\bold r,\omega) \neq 0$ in regions A or B.  Figure 1 shows the special case where dielectric bodies A \& B are parallel plates, but our expressions do not assume that geometry.

The partition function of the coupled dielectric system is
$\mathrm Z=\int \mathcal D \phi \, \exp\left\{i S\right\}$. Without the presence of dielectric materials A \& B, the partition function is $\mathrm Z_0$, which can be obtained from $\mathrm Z$ by setting $\chi^\prime_A=\chi^\prime_B=0$.
Formally, then, the energy $\rm E$ of the coupled systems A \& B can be obtained from the reduced partition function, yielding
\begin{eqnarray}\label{eq3}
\mathrm E=\frac{i}{\mathrm T}\mathrm{\ln\,\frac{Z}{Z_0}}=
\mathrm{\int\limits_0^{\infty} \frac{d\xi}{2\pi} \ln \,Det\left(1+ \hat{\chi}^{\prime}(\bold r, i \xi) \xi^2 \hat G_0\right)}.
\end{eqnarray}
Here the integral is evaluated in the complex frequency plane $\xi=-i\omega$. In this formula, $\rm T$ is the time interval in path integral formula, and $\rm \hat G_0(\xi)=(\chi_0(i\xi) \xi^2-\nabla^2)^{-1}$ is the Green's function for scalar field in medium C.  

Note that $\hat{\chi}^{\prime}$ is an operator, which takes different eigenvalues depending on its eigen-functions.  We divide the whole Hilbert space into three parts $\mathcal H=\mathcal H_A\oplus\mathcal H_B\oplus\mathcal H_C$, where $\mathcal H_{A,B,C}$ represent the Hilbert space for wave functions in materials A, B, C. Writing $\hat{\chi^{\prime}}|\psi_s\rangle=\chi_s^{\prime} |\psi_s\rangle$,  where $|\psi_s\rangle$ corresponds to the wave function in region $s$ (s=A, B, C), the energy of the coupling dielectric materials can be written in a matrix form
\begin{eqnarray}\label{eq4}
\rm E =&\nonumber \\
&\int\limits_0^{\infty} \frac{d\xi}{2\pi}\ln \mathrm{Det} \left(
\begin{array}{cc}
1_A+\chi_{\rm A}^{\prime} \xi^2{\rm U}_{\rm AA}  & \chi_{\rm A}^{\prime}\xi^2{\rm U}_{\rm AB} \\
\rm \chi_B^{\prime}\xi^2U_{BA} & \rm 1_B+\chi_B^{\prime}\xi^2U_{BB} 
\end{array}\right)
\end{eqnarray}
where $\rm U_{ss^{\prime}}=\langle \psi(x\in s)|\hat G_0|\psi(x\in s^{\prime})\rangle$ ($s$,$s^{\prime}$=$\rm A, B$) is the propagator between A and B.  The diagonal elements in Eqn. \eqref{eq4} correspond to the self-energy of material A and B, which is independent of their relative distance.
The Casimir energy $\rm E_c$ between A \& B (i.e., the coupling energy between A \& B), can be obtained by subtracting the diagonal parts of $\rm E$, yielding
\begin{eqnarray}\label{eq5}
\rm E_c=\int_0^{\infty}\frac{d\xi}{2\pi}\ln\, Det \left(
1-T_AU_{AB}T_BU_{BA}\right)
\end{eqnarray}
where $\rm T_s=\chi_{s}^{\prime}\xi^2/(1+\chi_{s}^{\prime}\xi^2 U_{ss})$ ($s= \rm A, B$). 
Eqn. \eqref{eq5} has a ready interpretation in terms of Feynman diagrams and conventional quantum electrodynamic perturbation theory \cite{Dzyaloshinskii}. In an isotropic medium, left-circular polarized and right-circular polarized photons are equivalent, so that photon Green's function can be represented by a single wavy line (figure 2(a)). 
\begin{figure}[!htb]
\centering
\includegraphics[height=2.8cm, width=6.8cm, angle=0]{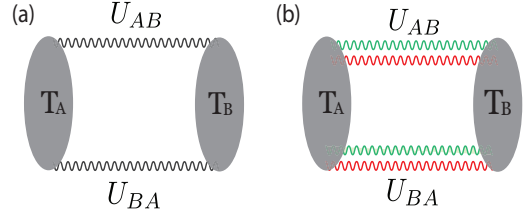}
\caption{Feynman diagrams for normal Casimir energy and chiral Casimir energy. Figure (a) shows the Feynman diagram representation for normal Casimir energy when chiral symmetry of photon is kept. Black wavy lines represent photon propagator $\rm \hat D_0$ and filled bubbles represent current-current correlation functions $\rm T_{A}$ \& $\rm T_B$. Figure (b) shows the Feynman diagram representation for Casimir energy when chiral symmetry is broken, namely, the velocity of photons depend on their chirality. Red and green wavy lines correspond to Green's functions for right-circular polarized photons and left-circular polarized photons, respectively.}\label{figure2}
\end{figure}

Now let us review the logic of the central no-go theorem \cite{oken}.  If there is reflection symmetry between A and B, then the self-energy operators $\rm T_A$ and $\rm T_B$ are related by a reflection operator $\mathcal J_m$, according to
$\rm T_B=\mathcal J_m^\dagger T_A \mathcal J_m$. Moreover, one can show that $\rm U_{AB} \mathcal J_m=\mathcal J_m^\dagger U_{BA}$ is a Hermitian operator. Thus the Casimir energy can be expressed as
\begin{eqnarray}\label{eq6}
\rm E_c=\int_0^{\infty}\frac{d\xi}{2\pi} \ln\, Det \left(
1-(\sqrt{T_A}U_{AB}\mathcal J_m\sqrt{T_A})^2\right).
\end{eqnarray}
The integrand has the functional form $g(x)= \ln\left[1-f(x)^2\right]$, leading to $g^\prime(x)=-2f(x)f^{\prime}(x)/(1-f^2(x))$, so that $f(x)>0$ and $f^{\prime}(x)<0$ imply $g^\prime(x) > 0$. Within our integrand $\mathrm{I}(l) \equiv \langle \psi|\mathrm{U_{AB}\mathcal J_m}|\psi\rangle>0$ and $\partial_l\mathrm{I}(l)<0$. Consequently, the Casimir force $\mathrm{F_c}=-d\mathrm{E_c}/dl<0$ between A and B is attractive. 

The foregoing procedures and arguments are readily adapted to the electromagnetic field case. In the gauge $A_0=0$ one has $S=\frac{1}{2}\int d\bold r \frac{d\omega}{2\pi}A_\omega^*\left[-\nabla\times\nabla\times+\epsilon(\bold r,\omega) \omega^2\right]A_\omega$, which matches the massless scalar field form. The electromagnetic Casimir energy for electromagnetic field results from substituting $\rm \hat G_0\mapsto \hat D_0(i\xi)=\left(\frac{1}{\nabla\times\nabla\times+\chi_0(i\xi)\xi^2}\right)$ in Eqn. \eqref{eq4}, and interpreting $\rm U_{ss^{\prime}}$ appropriately.  Thus the no-go theorem still applies.

The loophole appears when we note that in chiral media,  $\mathrm{U_{AB}} \mathcal J_m$ is not a Hermitian operator, i.e., $\rm U_{AB}\mathcal J_m \neq \mathcal J_m^{\dagger} U_{BA}$.   This arises, physically, because there are different phase velocities for  left-circular polarized photons versus right-circular polarized photons.  

To model chiral media, we assume a chirality-dependent dielectric function in material C,  i.e., $\rm \chi_0^{L(R)}$ for left- and right-circular polarized photons.   The Green's function must be written in a matrix form in chiral basis ($\psi_L(x),\psi_R(x)$), i.e.,  
\begin{eqnarray}\label{eq7}
\rm \hat D_0=\left(\begin{array}{cc}
\rm \hat D_0^{L}& 0\\
0& \rm \hat D_0^{R}
\end{array}\right)
\end{eqnarray}
where $\rm \hat D_0^{L(R)}=(\chi_0^{L(R)}(i\xi)\xi^2+\nabla\times\nabla\times)^{-1}$ represent the Green's function for left(right)-circular polarized photons. Figure 2(b) shows the Feynman diagram for chiral Casimir energy.  To keep track of the chiral degree of freedom, it is helpful to use a double wavy line to represent the photon Green's function. Even when the reflection symmetry is kept between A and B, through their identical properties and symmetric geometry, the material C breaks the symmetry.  
The propagators in the Feynman diagram exchange colors (red $\leftrightarrow$ green) under the reflection operation $\mathcal J_m$.  Now $\rm \mathcal J_m^{\dagger} U_{BA}\mathcal J_m =I_AU_{AB}I_A\neq U_{AB}$, where $\rm I_A$ is an off-diagonal unit matrix.  Thus, $\rm T_AU_{AB}T_{B}U_{BA} \neq (\sqrt{T_A}U_{AB}\mathcal J_m\sqrt{T_A})^2$, and the foregoing arguments fail.


\textit{Calculations for chiral media in plate geometry}: By using a non-reciprocal Green's function method, we can derive more tractable expressions for chiral Casimir forces.  The algebra, which is not entirely trivial, is set out in the supplemental materials \cite{supplement}. (Compare \cite{sma1,sma2,sma3,bordag1}.)

Specializing to  plate geometry, we find the energy per unit surface area 
\begin{eqnarray}\label{eq8}
\rm E_c=\int\limits_0^{\infty}\frac{d\xi}{2\pi} \int\limits_{-\infty}^{\infty} \frac{d^2k_{\|}}{(2\pi)^2} \left\{\ln\, Det \left(I-R_B\tilde U_{BA}R_A\tilde U_{AB}\right)\right\}
\end{eqnarray}
where $\xi$ is the imaginary frequency, and $\rm \bold k_{\|}=(k_x,k_y)$ represents momentum in $xy$ plane (parallel with plates). 
Here $\rm R_{A}$ ($\rm R_B$) represents reflection matrix at plate A (B), and $\rm \tilde U_{AB}$ ($\rm \tilde U_{BA}$) represents translation matrix from A to B (B to A).  (Note that this $\tilde U$ has quite a different meaning from $U$, which appeared earlier.)

In a chiral medium, reflection symmetry of photons is broken, implying that TE (s-polarized) wave and TM (p-polarized) wave are not the eigenstates. In the more convenient chiral basis, $\rm \tilde U_{AB}$ and $\rm \tilde U_{BA}$ are diagonal, as long as chirality itself is a good quantum number.  We have then simply 
\begin{eqnarray}\label{eq9}
\rm \tilde U_{BA}=\left(
\begin{array}{cc}
e^{i  k_{zL}^+l}&0\\
0&e^{ik_{zR}^+ l}
\end{array}\right),\,\,
\rm \tilde U_{AB}=\left(
\begin{array}{cc}
e^{i k_{zL}^{-}  l}&0\\
0&e^{i k_{zR}^{-}l}
\end{array}\right)
\end{eqnarray}
where $k_{zR}^{\pm}$ and $k_{zR}^{\pm}$ stand for the propagating velocity of right-circular polarized photons and left-circular polarized photons, respectively. The superscript $\pm$ indicates the propagating directions of photons. (The meaning of $k_{zR/L}^{\pm}$ is also shown in figure 1.)
However,  photons can change chirality at the boundary A \& B due to reflection.  In this paper, we only consider the case where there is reflection symmetry between  A \& B, implying the same reflection matrices of A \& B:
\begin{eqnarray}\label{eq10}
\rm R_{A}=R_{B}=\left(
\begin{array}{cc}
\rm r_{RR}&\rm r_{LR}\\
\rm r_{RL}&\rm r_{LL}
\end{array}\right),
\end{eqnarray}
where $\mathrm{r}_{ij}$ represent the reflection magnitude of a photon from chirality $j$ to $i$ ($i,j=\mathrm{L,R}$). 

Eqn. \eqref{eq8} can be interpreted integrating over round trips of virtual photons. First imagine that a photon goes from B to A with translation matrix ($\rm \tilde U_{AB}$), and then is reflected at plate A ($\rm R_A$). After its first reflection, it goes back from A to B with translation matrix ($\rm \tilde U_{BA}$), and then it will be reflected at plate B ($\rm R_B$) again. 

\begin{figure}[!htb]
\centering
\includegraphics[height=3.2cm, width=8.5cm, angle=0]{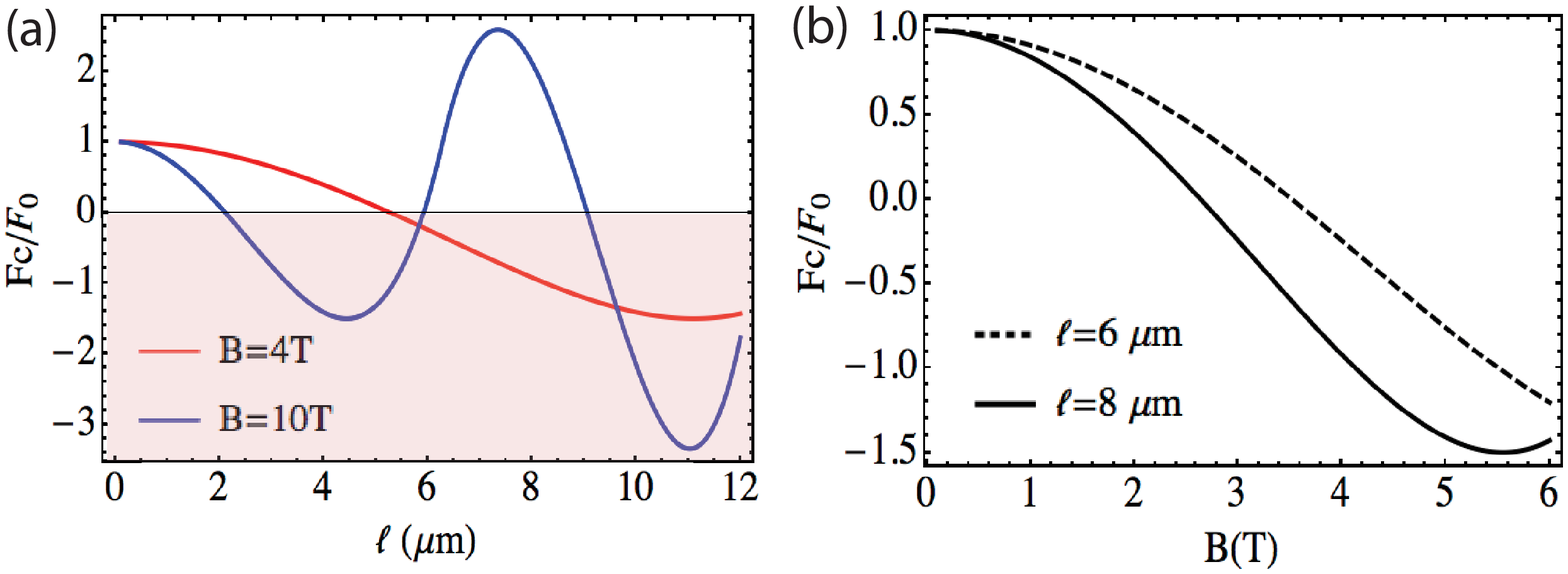}
\caption{Chiral Casimir force due to Faraday effect, normalized to the original metallic Casimir force. (a) shows the Casimir force enhancement in different magnetic field. The red and blue curves represent Casimir force at magnetic field $\rm B=4 \,T$ and $\rm B=10 \,\mathrm T$, respectively. The shadow region corresponds to repulsive Casimir force regime.  (b) shows how the magnetic field $\rm B$ can control the Casimir force. The solid line and dash line represent the Casimir force that is measured at the distance $l=8 \mathrm{\mu m}$ and $l=6 \mathrm{\mu m}$, respectively. \label{figure3}}
\end{figure}

(i) Faraday materials.  In a medium displaying the Faraday effect, the optical rotation angle $\theta$ is determined by $\theta=\mathcal V B l$, where $\mathcal V$ is the Verdet constant (a key parameter in Faraday materials), $B$ is the magnetic field in the light propagating direction, and $l$ is the distance that the light pass through.  In an alternate description, the magnetic field introduce a phase velocity difference $\delta k_z=\mathcal V B$ between left-circular polarized photons and right-circular polarized photons. Therefore, the wave vectors of photons with different chirality satisfy  $k_{zR}^{+}=k_{zL}^{-}=\bar k_z+ \delta k_z$ and $k_{zR}^{-}=k_{zL}^{+}=\bar k_z - \delta k_z$, where $\bar k_z$ is the average wave vector of right-circular and left-circular polarized photons \cite{footnote2}. With the phase velocity expressions of chiral photons, one can obtain the translation matrices $\rm \tilde U_{AB}$ ($\rm \tilde U_{BA}$) for Faraday materials.  For ideal metal plates, the reflection matrices are simply taken off-diagonal unit matrices, i.e., $\rm r_{RR}=r_{LL}=0$ and $\rm r_{LR}=r_{RL}=-1$.
The off-diagonal reflection matrix is due to the fact that photons change their chirality after being reflected at an ideal metal plate \cite{footnote3}.   

Recently, experiments have measured very large Verdet constants in some organic molecules and liquids.  We set Verdet constant as $\mathcal V=5\times10^4 \mathrm{\,rad \,m^{-1}\,T^{-1}}$ in the calculation based on several experimental results \cite{crassee}. (We will consider frequency-dependent Verdet constant further below.)
Substituting the expression of reflection matrices and translation matrices into Eqn. \eqref{eq7}, one obtains the Casimir energy for Faraday materials
\begin{eqnarray}\label{eq11}
\mathrm{E_c=\int\limits_0^{\infty} \frac{d\xi}{2\pi}\int\limits_{-\infty}^{\infty}\frac{d^2 k_{\|}}{(2\pi)^2}  \ln} \left[ (1+e^{-4\kappa l}-2e^{-2\kappa l}\mathrm{cos}(2\mathcal V B l)\right] &{}&\nonumber \\
&{}&
\end{eqnarray}
where $\kappa=\sqrt{\xi^2+k_{\|}^2}=\sqrt{\xi^2+k_x^2+k_y^2}$. From Eqn. \eqref{eq11}, one finds that magnetic field and Verdet constant are embedded within the expression of Casimir energy. Therefore, the Casimir force can be manipulated by tuning magnetic field. Moreover, the cosine function in the expression leads to the repulsive Casimir force. 
Figure 3 shows the repulsion and enhancement of Casimir force in different magnetic field. $\rm F_0$ represents the Casimir force with Verdet constant $\mathcal V=0$, i.e., no gyrotropic materials inserted between A \& B. In contrast, $\rm F_c$ represents the gyrotropic Casimir force. The ratio $\rm F_c/F_0<0$ being negative indicates the emergence of repulsive Casimir force, whereas the ratio $\rm |F_c/F_0|>1$ indicates the enhancement of Casimir force.

\begin{figure}[!htb]
\centering
\includegraphics[height=3.8cm, width=5.8cm, angle=0]{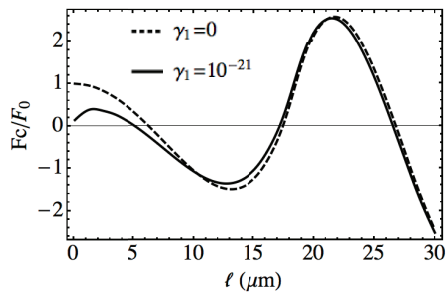}
\caption{Chiral Casimir force between two parallel plates mediated by an optically active material. Solid (dashed) curve corresponds to the ratio of Casimir force $\rm F_{c}/F_{0}$ for frequency-dependent(-independent) specific rotation $\alpha_0$.  Based on experimental results in optically active materials \cite{emile},  the parameters are chosen as: $\omega_p=10^{16} \rm s^{-1}$, $\gamma_0=2\times10^{6}\rm deg\,dm^{-1}g^{-1}cm^{3}$, $\gamma_1\approx10^{-21}\rm deg\,dm^{-1}g^{-1}cm^{3}s^2$ and $\rho=1 \rm g \,cm^{-3}$ in the calculation. \label{figure4}}
\end{figure}

(ii) Optically active materials.  In optically active medium,  the optical rotation angle has a similar form as that in Faraday medium, i.e.,  $\theta=\alpha_0 \rho\, l$, where $\alpha_0$ is called specific rotation (an important parameter in optically active materials), $\rho$ is mass concentration, and $l$ is the light propagating distance. Therefore, one can build a rough correspondence between Faraday effect and optically active effect via substitution
$\mathcal V B \mapsto \alpha_0\rho$. However, there are relevant differences between Faraday and optically active materials. In Faraday materials, the external magnetic field breaks time reversal symmetry, which, by contrast,  is preserved in optically active medium. Due to this difference, in an optically active medium the phase velocity of chiral photons does not depend on the direction of propagation; furthermore, if time reversal symmetry also holds in the reflection process, then the reflection matrices do not change chirality \cite{plum}. 
Without loss of generality, we assume that the velocity difference between opposite chirality of photons is $\delta k_z=\alpha_0 \rho$. Thus, the phase velocity of chiral photons in optically active materials satisfies $k_{zR}^{+}=k_{zR}^{-}=\bar k_z+ \delta k_z$ and $k_{zL}^{+}=k_{zL}^{-}=\bar k_z - \delta k_z$.   An especially simple result emerges if one chooses diagonal reflection matrices for A \& B, i.e.,  $\rm r_{RR}=r_{LL}=1$ and $\rm r_{RL}=r_{LR}=0$. In this case, the Casimir energy in chiral active material can be obtained by a substitution: $\mathcal V B \mapsto \alpha_0\rho$ in Eq.\eqref{eq11}.

Of course, one can also apply magnetic fields to optically active materials, resulting in more control, but more complex formulas. 

In general, the Verdet constant depends on the wavelength of light $\lambda$, which is usually modeled as $\mathcal V(\lambda)=a_0+b_0/\lambda^2$, where $a_0$ and $b_0$ are fitting parameters according to experimental results \cite{vandendriessche2}.  Notice that in the low frequency limit, $\mathcal V=a_0$, is most important for long range force.

In any real experiments, the distance between two bodies A \& B is always finite. Therefore, to obtain more accurate results, one needs take frequency dependence of effective Verdet constant $\mathcal V$ (or specific rotation $\alpha_0$ ) into consideration.  Moreover, photons are not perfectly reflected at real metal plates. We model the reflection coefficients as $\rm r_{RR}=r_{LL}=0$ and $\rm r_{RL}=r_{LR}=e^{-\omega/\omega_p}$ where $\rm \omega_p$ represents cutoff frequency. That means perfect reflection only happens in the low frequency. i.e., $\rm \omega<<\omega_p$. 
In figure 4, we examine the Casimir force mediated by an optically active material, where the frequency-dependent of specific rotation is modeled as $\alpha_0(i\xi)=\gamma_0+\gamma_1 \xi^2$. (This model is also often used in optically active materials \cite{vandendriessche2}.)  We see that the short distance behavior is modified quantitatively, but the long range limit is not influenced.

\textit{Summary}: We have identified an important loophole of the no-go theorem on Casimir force, and demonstrated that repulsive Casimir force could emerge between two similar bodies with reflection symmetry. The key to realizing repulsive Casimir forces  between similar objects is to insert an intermediate chiral material between them. The chiral Casimir force has several distinctive features: it can be oscillatory, its magnitude can be large (relative to the classic Casimir force), and it can vary in response to external magnetic fields.  Through the connection of this force to independently measurable material properties, one obtains a wealth of predicted phenomena which directly reflect macroscopic effects of quantum fluctuations.  Finally, let us call attention to the finite temperature extension of these results \cite{supplement}, which brings in larger forces with similar qualitative behavior. 

\bigskip

\textit{Acknowledgement}: This work was supported by the Swedish Research Council under Contract No. 335-2014-7424.  In addition, FW's work is supported by the U.S. Department and by the European Research Council under grant 742104.


\hspace{3mm}


\clearpage
\begin{widetext}

\begin{center}
\textbf{Supplemental Materials for ``Chiral Casimir Forces: Repulsive, Enhanced, Tunable''}
\end{center}

\begin{center}
Qing-Dong Jiang  and  Frank Wilczek
\end{center}
\maketitle

\vspace{2mm}

\begin{center}
\textbf{${\rm\uppercase\expandafter{\romannumeral1}}$. Derivation of gyrotropic Casimir force by non-reciprocal Green's function.}
\end{center}

We will derive the chiral Casimir energy formula Eqn. \eqref{eq8} in the main text with the aid of non-reciprocal Green's function tensor $\mathds G(\bold r^{\prime},\bold r,\omega)$. 
\\

(i) \textit{Tensor expression of Casimir force.}  

First, let us recall the Maxwell equations in a material:
\begin{eqnarray}
\nabla\cdot \bold E&=&\rho/\epsilon_0\\
\nabla\cdot\bold B&=&0\\
\nabla\times\bold E&=&-\frac{\partial \bold B}{\partial t}\\
\nabla\times\bold B&=&\mu_0\left(\bold j+\epsilon_0\frac{\partial \bold E}{\partial t}\right),
\end{eqnarray}
where $\bold E$ and $\bold B$ represent electric and magnetic induction, whereas $\rho$ and $\bold j$ correspond to the total charge and current density. The Lorentz force density is given by
\begin{eqnarray}
\rm \bold f=\rho \bold E+\bold j\times \bold B.
\end{eqnarray}
With the help of Maxwell equations, the Lorentz force density can be re-written as \cite{raabe}
\begin{eqnarray}
\rm \bold f=\nabla\cdot \mathds T-\epsilon_0 \frac{\partial}{\partial t}(\bold E\times\bold B),
\end{eqnarray}
where the stress tensor $\mathds T$ is expressed by electric field and magnetic induction, i.e.,
\begin{eqnarray}
\rm \mathds T=\epsilon_0\bold E\otimes \bold E+\mu_0^{-1}\bold B\otimes\bold B-\frac{1}{2}(\epsilon_0 \bold E\cdot\bold E+\mu_0^{-1}\bold B\cdot\bold B)\mathds I
\end{eqnarray}
In this equation, $\otimes$ represents the tensor product, and $\mathds I$ represents the unit tensor. Notice that the term $\rm \frac{\partial}{\partial t}(\bold E\times\bold B)$ actually means the change of Poynting vector in time. In order to obtain the total force, one must integrate the Lorentz force density over the whole considered region. The Casimir force can be obtained by evaluating the vacuum state average of the stress tensor, i.e. $\langle \mathds T\rangle$.  Thus the Casimir force between two bodies with total surface area $\bold A$ can be  written as
\begin{eqnarray}
\bold F_c=\int_{\partial V}d\bold A\cdot\langle \epsilon_0 \bold E(\bold r)\otimes\bold E(\bold r^{\prime})+\frac{1}{\mu_0}\bold B(\bold r)\otimes\bold B(\bold r^{\prime})-\frac{1}{2}\left[\epsilon_0 \bold E(\bold r)\cdot\bold E(\bold r^{\prime})+\frac{1}{\mu_0}\bold B(\bold r)\cdot\bold B(\bold r^{\prime})\right] \mathds I\rangle_{\bold r^{\prime}\rightarrow \bold r},
\end{eqnarray}
where the subscript  $\bold r^{\prime}\rightarrow \bold r$ implies that one needs take the limit $\bold r^{\prime}=\bold r$ in the final expression in order to remove the self-energy in the cavity \cite{raabe}. 
\\

(ii)\textit{Green's function expression of Casimir force.}

Due to fluctuation-dissipation theorem (FDT), the expectation value of electric and magnetic field can be obtained from Green's tensor \cite{raabe}:
\begin{eqnarray}
\rm \langle \bold E(\bold r)\otimes\bold E(\bold r^{\prime})\rangle&=&\rm \frac{\hbar \mu_0}{\pi}\int_{0}^{\infty} d\omega \omega^2 Im\left\{\mathds{G}(\bold r,\bold r^{\prime},\omega)\right\}\\
\rm \langle \bold B(\bold r)\otimes\bold B(\bold r^{\prime})\rangle&=&\rm -\frac{\hbar \mu_0}{\pi}\int_{0}^{\infty} d\omega \overrightarrow{\nabla}\times Im\left\{\mathds{G}(\bold r,\bold r^{\prime},\omega)\right\}\times \overleftarrow{\nabla}^{\prime}.
\end{eqnarray}
However, due to the presence of chiral material, Lorentz reciprocity is violated, which means $\mathds G^{T}(\bold r^{\prime},\bold r,\omega)\neq \mathds G(\bold r,\bold r^{\prime},\omega)$. Therefore, one have to redefine the real and imaginary parts of Green's tensor  \cite{buhmann}:
\begin{eqnarray}
\rm Re\{\mathds{G}(\bold r, \bold r^{\prime},\omega)\}&=&\frac{1}{2} \left[\mathds{G}(\bold r, \bold r^{\prime},\omega)+\mathds{G}^{*T}(\bold r^{\prime},\bold r, \omega)\right]\\
\rm Im\{\mathds{G}(\bold r, \bold r^{\prime},\omega)\}&=&\frac{1}{2i} \left[\mathds{G}(\bold r, \bold r^{\prime},\omega)-\mathds{G}^{*T}(\bold r^{\prime},\bold r, \omega)\right].
\end{eqnarray}
Substituting the electromagnetic field operators with non-reciprocal Green's tensor, one obtains the Casimir force in terms of non-reciprocal Green's function, i.e.,
\begin{eqnarray}
\begin{split}
\rm \bold F_c=-\frac{\hbar}{2\pi}\int_{0}^{\infty} d\xi \int_{\partial V}d\bold A\cdot &\left\{\frac{\xi^2}{c^2}\mathds G(\bold r,\bold r^{\prime},i\xi)+\frac{\xi^2}{c^2}\mathds G^{T}(\bold r^{\prime},\bold r,i\xi)+\overrightarrow{\nabla}\times \mathds{G}(\bold r,\bold r^{\prime},i\xi)\times \overleftarrow{\nabla}^{\prime}+\overrightarrow{\nabla}\times \mathds{G}^{T}(\bold r^{\prime},\bold r,i\xi)\times \overleftarrow{\nabla}^{\prime}\right.\\-&\mathrm{Tr}\left.\left[\frac{\xi^2}{c^2}\mathds G(\bold r,\bold r^{\prime},i\xi)+\overrightarrow{\nabla}\times \mathds{G}(\bold r,\bold r^{\prime},i\xi)\times \overleftarrow{\nabla}^{\prime}\right]\mathds I\right\}_{\bold r^{\prime}\rightarrow \bold r}.
\end{split}
\end{eqnarray}
In this formula, $\xi=-i\omega$ is the imaginary frequency, the same as in the main text,  This expression for Casimir force appears in ref. \cite{sfuchs}.  It is, however, not trivial to evaluate it. 

In the following section, we obtain the non-reciprocal Green's tensor $\mathds G(\bold r, \bold r^{\prime},i\xi)$ with the help of reflection matrices $\mathcal R_{\pm}$ and translation matrices $\rm U_{\pm}$ as is shown in figure \ref{figure5}.
\\

\begin{figure*}[!htb]
\centering
\includegraphics[height=4cm, width=10cm, angle=0]{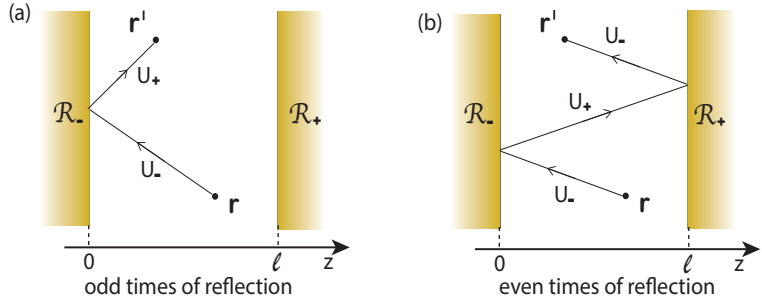}
\caption{Schematic illustration of scattering process between two plates with a gyrotropic medium inserted. (a) shows odd times of reflection, where the first reflection takes place at the left plate. For simplicity, we only show one time reflection case. (b) shows even times of reflection, where the first reflection takes place at the right plate. Again, we only show the least two times of reflection case for simplicity. \label{figure5}}
\end{figure*}

(iii) \textit{Derivation of the non-reciprocal Green's function in a cavity made by two parallel plates.}

With gauge fixing conditions, any polarization of photons can be decomposed into perpendicular polarization (TE wave or s polarization) and parallel polarization (TM wave or p polarization). We define the unit vector in s- and p-polarization directions as:
\begin{eqnarray}\label{esep}
e_{s\pm}&=&e_{k_{\|}}\times e_z=\frac{1}{k_{\|}}(k_y,-k_x,0)\\
e_{p\pm}&=&\frac{1}{k}(k_{\|}e_z\mp k_z e_{k_{\|}})=\frac{1}{k}\left(\mp \frac{k_z}{k_{\|}}k_x,\mp \frac{k_z}{k_{\|}}k_y,k_{\|}\right),
\end{eqnarray}
where $ k^2=k_{\|}^2+k_z^2=k_x^2+k_y^2+k_z^2$. The indices $\pm$ in the unit vectors in Eqn. \eqref{esep} refer to the propagating direction of photons. $+$ sign means that photons propagate to the right, whereas $-$ sign means that photons propagate to the left.
With the presence of chiral material, chirality of photons is a good quantum number, but not parity. Thus, it is convenient to switch to a chiral basis, defined as
\begin{eqnarray}
e_{R\pm}&=&\frac{1}{\sqrt{2}}(e_{p\pm}+ie_{s\pm});\\
e_{L\pm}&=&\frac{1}{\sqrt{2}}(e_{p\pm}-ie_{s\pm}).
\end{eqnarray}
It should be noted that $(e_{R +},e_{R -})^{T}\cdot(e_{L+}, e_{L-})=(e_{L+}, e_{L-})^{T}\cdot (e_{R+}, e_{R-})=\mathrm I$ and $(e_{R +},e_{R -})^{T}\cdot(e_{R+}, e_{R-})=(e_{L+}, e_{L-})^{T}\cdot (e_{L+}, e_{L-})=\mathrm I_A$, where $\mathrm I$ represents unit matrix as before, whereas $\mathrm I_A$ represents off-diagonal unit matrix, i.e.,
\begin{eqnarray}
\mathrm I_A=\left(\begin{array}{cc}
0&1\\1&0
\end{array}\right).
\end{eqnarray}
We will frequently use the matrix $\rm I_A$ in the derivation of chiral Casimir force.

In the following, we use $\mathds R$ and $\mathds U$ to represent reflection and translation tensor, respectively.  
The Green's tensor can be expressed by the combination of $\mathds R$ and $\mathds U$ . We show how to do this in the following. There are four distinguishable paths that contribute to the Green's tensor, i.e., odd or even times of reflection, and the first reflection happens at left or right boundary. In the figure \ref{figure5}, we only show two cases: the odd and even times of reflection with the first reflection occurring at left boundary. It is also easy to show the case where the first reflection occurs at the right boundary. We then define reflection and translation matrices $\mathrm R$ and $\mathrm U$ as:
\begin{eqnarray}
\mathds R_{\pm}&=&(e_{R\mp}, e_{L\mp}) \mathrm R_{\pm}\left(\begin{array}{cc}e_{R\pm}\\e_{L\pm}\end{array}\right)\\
\mathds U_{\pm}&=&(e_{R\pm}, e_{L\pm}) \mathrm U_{\pm}\left(\begin{array}{cc}e_{R\pm}\\e_{L\pm}\end{array}\right),
\end{eqnarray}
where $\mathds R_{\pm}$ ($\rm R_{\pm}$) represents reflection tensor (matrix) at right and left boundaries. Similarly, $\mathds U_{\pm}$ ($\rm U_{\pm}$) stands for translation tensor (matrix) goes to right and left, respectively.  With translational invariance, we demand $\mathds U_{\pm}(l-z)\mathds U_{\pm}(z)=\mathds U_{\pm}(l)$ for $z\in [0,l]$, where $l$ is the distance between left plate and right plate (see figure \ref{figure5}).  This condition indicates the equality $\mathrm U(l-z)\mathrm I_A\mathrm U(z)=\mathrm U(l)$.  Assume that a plane wave starts from a source point located at $\bold r$, and we want to obtain the field at point $\bold r^{\prime}$ (see figure \ref{figure5}). There are four possible reflection configurations in total:\\\\
(i) odd times of reflection, where the first reflection occurs at left boundary\\
$\mathds A=\mathds U_+(z)\mathds R_-\mathds U_-(z)+\mathds U_+(z)\mathds R_-\mathds U_-(l)\mathds R_+\mathds U_+(l)\mathds R_-\mathds U_-(z)+...$;\\
(ii) odd times of reflection, where the first reflection occurs at right boundary\\
$\mathds B=\mathds U_-(l-z)\mathds R_+\mathds U_+(l-z)+\mathds U_-(l-z)\mathds R_+\mathds U_+(l)\mathds R_-\mathds U_-(l)\mathds R_+\mathds U_+(l-z)+...$;\\
(iii) even times of reflection, where the first reflection occurs at right boundary\\
$\mathds C=\mathds U_+(z)\mathds R_-\mathds U_-(l)\mathds R_+\mathds U_+(l-z)+\mathds U_+(z)\mathds R_-\mathds U_-(l)\mathds R_+\mathds U_+(l)\mathds R_-\mathds U_-(l)\mathds R_+\mathds U_+(l-z)+...$;\\
(iv) even times of reflection, where the first reflection occurs at left boundary\\
$\mathds D=\mathds U_-(l-z)\mathds R_+\mathds U_+(l)\mathds R_-\mathds U_-(z)+\mathds U_-(l-z)\mathds R_+\mathds U_+(l)\mathds R_-\mathds U_-(l)\mathds R_+\mathds U_+(l)\mathds R_-\mathds U_-(z)+...$.\\\\
In the above expressions, $\mathds A$, $\mathds B$, $\mathds C$, and $\mathds D$ are four tensors represent four different, possible contribution to the Green's tensor. For simplicity, we introduce $\rm A$, $\rm B$, $\rm C$, and $\rm D$ as matrices which are defined by their corresponding tensors in the same way as $\rm R$ and $\rm U$ from tensors $\mathds R$ and $\mathds U$.
We are now ready to deal with chiral materials, where chirality of photons is good quantum number. However, the velocity of photons depend on chirality. Here, the wave vector of right-circular (left-circular) polarized photons in z-direction is assumed to be $k_{zR}^{\pm}$ ($k_{zL}^{\pm}$). $\pm$ in the upper indices denotes the propagating direction of photons.  In optically active materials, the velocities satisfy $k_{zR/L}^{\pm}=k_{zR/L}^{\mp}$, whereas, the z-direction wave vectors satisfy $k_{zR}^{\pm}=k_{zL}^{\mp}$ in Faraday materials [One can refer Table I to find out the phase velocities in Faraday materials or optically active materials.].  

To begin with, we deal with optically active materials first, and the results can be easily extended to Faraday materials with properly redefining chiral basis. In optically active materials, we assume the wave vector in z-direction to be $k_{zR}^{\pm}=\bar k_z+\delta k_z$ and $k_{zL}^{\pm}=\bar k_z-\delta k_z$ for right-circular and left-circular polarized photons, respectively. Left-circular polarized photons are a little faster than right-circular polarized photons.
With chiral basis, the Green's tensor can be written as
\begin{eqnarray}
\begin{split}
\mathds G(\bold r,\bold r^{\prime},\omega)=\frac{i}{8\pi^2}\mathrm{\int d^2 k_{\|} e^{i \bold k_{\|}\cdot(\bold r^{\prime}-\bold r)}}&\mathrm{\left\{
(e_{R+},e_{L+}) \mathrm A \left(\begin{array}{cc}e_{R-}/k_{zR}^{-}\\e_{L-}/k_{zL}^{-}\end{array}\right)
+(e_{R-},e_{L-}) \mathrm B \left(\begin{array}{cc}e_{R+}/k_{zR}^{+}\\e_{L+}/k_{zL}^{+}\end{array}\right)\right.}+\\
&\mathrm{\left.  (e_{R+},e_{L+}) \mathrm C \left(\begin{array}{cc}e_{R+}/k_{zR}^{+}\\e_{L+}/k_{zL}^{+}\end{array}\right)
+(e_{R-},e_{L-}) \mathrm D \left(\begin{array}{cc}e_{R-}/k_{zR}^{-}\\e_{L-}/k_{zL}^{-}\end{array}\right)
\right\}}.
\end{split}
\end{eqnarray}
With rotational symmetry, the integral can be easily evaluated in polar coordinate by using the substitution $\rm \int d^2 \bold k_{\|}=\int_0^{\infty} d k_{\|} k_{\|}\int_0^{2\pi} d\phi$, where $\rm k_{\|}=\sqrt{k_x^2+k_y^2}$ represents the absolute value of wave vector in xy plane, and $\phi$ represents the polar angle. To derive the Green's tensor above, one also needs verify different combinations of chiral basis, which we list as following:
\begin{eqnarray}
\rm{
\begin{split}
\int_{0}^{2\pi}d\phi\,\, e_{R\pm}e_{R\pm}&=\int_{0}^{2\pi}d\phi(e_{p\pm}+ie_{s\pm})(e_{p\pm}+ie_{s\pm})=\pi \left\{\left(\frac{{k_{zR}^{\pm}}^2}{k^2}-1\right)(e_xe_x+e_ye_y)+2\frac{k_{\|}^2}{k^2}e_ze_z\right\};\\
\int_{0}^{2\pi}d\phi\,\, e_{L\pm}e_{L\pm}&=\int_{0}^{2\pi}d\phi (e_{p\pm}-ie_{s\pm})(e_{p\pm}-ie_{s\pm})=\pi \left\{\left(\frac{{k_{zL}^{\pm}}^2}{k^2}-1\right)(e_xe_x+e_ye_y)+2\frac{k_{\|}^2}{k^2}e_ze_z\right\};\\
\int_{0}^{2\pi}d\phi\,\, e_{R\pm}e_{R\mp}&=\int_{0}^{2\pi}d\phi (e_{p\pm}+ie_{s\pm})(e_{p\mp}+ie_{s\mp})=\pi \left\{  \left(-\frac{k_{zR}^{+}k_{zR}^{-}}{k^2}-1\right)(e_xe_x+e_ye_y)+2\frac{k_{\|}^2}{k^2}e_ze_z\right\};\\
\int_{0}^{2\pi}d\phi\,\,e_{L\pm}e_{L\mp}&=\int_{0}^{2\pi}d\phi (e_{p\pm}-ie_{s\pm})(e_{p\mp}-ie_{s\mp})=\pi  \left\{ \left(-\frac{k_{zL}^{+}k_{zL}^{-}}{k^2}-1\right)(e_xe_x+e_ye_y)+2\frac{k_{\|}^2}{k^2}e_ze_z\right\};\\
\int_{0}^{2\pi}d\phi\,\,e_{R\pm}e_{L\mp}&=\int_{0}^{2\pi}d\phi (e_{p\pm}+ie_{s\pm})(e_{p\mp}-ie_{s\mp})=\pi \left\{ \left(-\frac{k_{zR}^{\pm}k_{zL}^{\mp}}{k^2}+1\right)(e_xe_x+e_ye_y)+2\frac{k_{\|}^2}{k^2}e_ze_z\right\};\\
\int_{0}^{2\pi}d\phi\,\,e_{L\pm}e_{R\mp}&=\int_{0}^{2\pi}d\phi (e_{p\pm}-ie_{s\pm})(e_{p\mp}+ie_{s\mp})=\pi \left\{ \left(-\frac{k_{zL}^{\pm}k_{zR}^{\mp}}{k^2}+1\right)(e_xe_x+e_ye_y)+2\frac{k_{\|}^2}{k^2}e_ze_z\right\};\\
\int_{0}^{2\pi}d\phi\,\,e_{R\pm}e_{L\pm}&=\int_{0}^{2\pi}d\phi (e_{p\pm}+ie_{s\pm})(e_{p\pm}-ie_{s\pm})=\pi \left\{ \left(\frac{k_{zR}^{\pm}k_{zL}^{\pm}}{k^2}+1\right)(e_xe_x+e_ye_y)+2\frac{k_{\|}^2}{k^2}e_ze_z\right\};\\
\int_{0}^{2\pi}d\phi\,\,e_{L\pm}e_{R\pm}&=\int_{0}^{2\pi}d\phi (e_{p\pm}-ie_{s\pm})(e_{p\pm}+ie_{s\pm})=\pi \left\{ \left(\frac{k_{zL}^{\pm}k_{zR}^{\pm}}{k^2}+1\right)(e_xe_x+e_ye_y)+2\frac{k_{\|}^2}{k^2}e_ze_z\right\}.
\end{split}}
\end{eqnarray}
In order to calculate the Green's tensor $\overrightarrow{\nabla}\times \mathds{G}(\bold r,\bold r^{\prime},\omega)\times \overleftarrow{\nabla}^{\prime}$, one has to apply operators $\overrightarrow \nabla$ and $\overrightarrow \nabla$ to chiral basis from the left and the right, respectively.  In momentum space representation, one is allowed to make the substitution, \textit{viz},  $\overrightarrow \nabla\times\rightarrow i\bold k_{\pm}\times$ and $\times\overleftarrow{\nabla}\rightarrow \times i\bold k_{\pm}$. Therefore, $e_R$ and $e_L$ transform as
$i\bold k_{\pm}\times e_{R\pm}=\omega e_{R\pm}$ and $i\bold k_{\pm}\times e_{L\pm}=- \omega e_{L\pm}$. With above preparation, one can readily obtain
\begin{eqnarray}
\rm{
\begin{split}
&\overrightarrow{\nabla}\times\mathds G(\bold r,\bold r^{\prime},\omega)\times\overleftarrow{\nabla}\\ &=-\frac{i\omega^2}{8\pi^2}\int d^2 k_{\|} e^{i \bold k_{\|}\cdot(\bold r^{\prime}-\bold r)}\left\{
(e_{R+},-e_{L+}) \mathrm A \left(\begin{array}{cc} e_{R-}/k_{zR}^{-}\\-e_{L-}/k_{zL}^{-}\end{array}\right)
+(e_{R-},-e_{L-}) \mathrm B \left(\begin{array}{cc} e_{R+}/k_{zR}^{+}\\-e_{L+}/k_{zL}^{+}\end{array}\right)\right.\\
&\qquad\qquad\qquad\qquad\qquad\left.  (e_{R+},-e_{L+}) \mathrm C \left(\begin{array}{cc}e_{R+}/k_{zR}^{+}\\-e_{L+}/k_{zL}^{+}\end{array}\right)
+(e_{R-},-e_{L-}) \mathrm D \left(\begin{array}{cc} e_{R-}/k_{zR}^{-}\\ -e_{L-}/k_{zL}^{-}\end{array}\right)
\right\}.
\end{split}}
\end{eqnarray}
To obtain explicit analytical results, let us consider the Casimir force between two parallel, uncharged, infinite plates. Due to rotational symmetry of the system, the Casimir force must be perpendicular to the surface.  In our case $d \bold A \| e_z$, and we only consider Casimir force in the z direction, which indicates that one only needs keep track of diagonal terms of $\mathds G$ and $\overrightarrow{\nabla}\times \mathds{G}(\bold r,\bold r^{\prime},\omega)\times \overleftarrow{\nabla}^{\prime}$ in calculation.  Substitute Green's tensor into the expression of Casimir force, and one can obtain
\begin{eqnarray}
\mathrm F_c=\frac{\hbar}{4\pi^2}\mathrm{\int_0^{\infty}d\omega\int_{0}^{\infty} d k_{\|} k_{\|}\left\{ \mathrm{OTr}\left\{\mathrm K \mathrm C\right\}+\mathrm{OTr}\left\{\mathrm K \mathrm D\right\}\right\}},
\end{eqnarray}
where $\rm OTr\{\}$ stands for anti-diagonal trace, which is defined as the summation of off-diagonal elements. For example, $\rm OTr\{S\}=S_{12}+S_{21}$ for a $2\times2$ matrix $\rm S$. In above formula, $\mathrm K$ is a diagonal matrix, where $\mathrm K_{11}=\bar k_z+\delta k_z$, $\mathrm K_{22}=\bar k_z-\delta k_z$, and $\rm K_{12}=K_{21}=0$. 
The diagonal elements  $\rm K_{11}$ and $\rm K_{22}$ stand for the wave vectors of right-handed and left-handed photons, respectively. Notably, for the convenience of the following context, we use real frequency representation of the Casimir force here.
Based on the tensor expressions of $\mathds C$ and $\mathds D$, the corresponding matrices $\rm C$ and $\rm D$ can be expressed by the reflection matrices $\mathrm R_\pm$ and the translation matrices $\rm U_{\pm}$, i.e.,
\begin{eqnarray}
\begin{split}
\mathrm {C}&= \mathrm{U_-}(l-z) \mathrm{I_A R_+ I_A U_+ I_A \left[I+ R_- I_A U_- I_A R_+ I_A U_+I_A +...\right]R_- I_A U_-}(z)\\
&=\mathrm{I_A \tilde{U}_-}(l-z)  \mathrm{R_+ \tilde{U}_+ \left[I+ R_- \tilde{U}_-  R_+ \tilde{U}_++...\right]R_-}\mathrm{ \tilde{U}_-}(z)\mathrm I_A
\end{split}
\end{eqnarray}
and
\begin{eqnarray}
\begin{split}
\mathrm D&=\mathrm{U_+}(z) \mathrm{I_A R_- I_A U_- I_A  \left[I+ R_+I_A U_+I_A R_- I_A U_-I_A+...\right]R_+I_AU_+}(l-z)\\
&=\mathrm{I_A \tilde{U}_+}(z) \mathrm{R_- \tilde{U}_-  \left[I+ R_+\tilde{U}_+ R_- \tilde{U}_-+...\right]R_+\tilde{U}_+}(l-z)\mathrm{I_A},
\end{split}
\end{eqnarray}
where $\rm I_A$ is an anti-diagonal unit matrix due to the fact that $\rm I_A=(\bold e_{R\pm}, \bold e_{L\pm})^T\cdot (\bold e_{R\pm},\bold e_{L\pm})$. The new translation matrix $\rm \tilde U$ is defined as $\rm \tilde U_{\pm}=I_A U_{\pm} I_A$. With some algebra on matrix and trace manipulation, one can show that $\rm OTr\{KC\}=Tr\{\tilde K M_C(I-M_C)^{-1} \}$ and $\rm OTr\{KD\}=Tr\{\tilde K M_D(I-M_D)^{-1} \}$, where 
\begin{eqnarray}
\mathrm{M_C}&=&\mathrm{R_+\tilde U_+R_-\tilde U_-}\\
\mathrm{M_D}&=&\mathrm{R_-\tilde U_-R_+\tilde U_+}
\end{eqnarray}
and the new $\rm \tilde K$ matrix is 
\begin{eqnarray}
\tilde K=\left(\begin{array}{cc}k_z-\delta k_z&0\\0&k_z+\delta k_z\end{array}\right).
\end{eqnarray}
Because $\rm U_{\pm}$ are diagonal matrices, one can verify the following identity:
\begin{eqnarray}
\begin{split}
-i\frac{\partial}{\partial l} \mathrm{Tr\left\{R_{+}\tilde U_{+}R_{-}\tilde U_{-}\right\} }&=-i \mathrm{Tr}\left\{\mathrm{R_+\tilde U_+R_-}\frac{\partial}{\partial l} \mathrm{\tilde U_-}\right\}-i\mathrm{Tr}\left\{\mathrm{R_-\tilde U_-R_+}\frac{\partial}{\partial l}\mathrm{\tilde U_+}\right\}\\ &=\rm Tr\left\{\tilde K R_+\tilde U_+R_-\tilde U_-\}+Tr\{\tilde K R_-\tilde U_-R_+\tilde U_+\right\}.
\end{split}
\end{eqnarray}
Then, one can re-express the Casimir force as 
\begin{eqnarray}
\mathrm {F_c}=-\frac{i\hbar}{4\pi^2}\mathrm{\int_0^{\infty}d\omega \int_0^{\infty} dk_{\|} k_{\|}Tr}\,\left\{\frac{\partial}{\partial l}\mathrm{\left(R_{+}\tilde U_{+}R_{-}\tilde U_{-}\right)\left(I-R_{+}\tilde U_{+}R_{-}\tilde U_{-}\right)^{-1}}\right\}.
\end{eqnarray}
From the expression of the chiral Casimir force, we find the chiral Casimir energy $E_c$ using identity $\rm Tr \ln \hat O=\ln \,Det \hat O$. The final Casimir energy is
\begin{eqnarray}\label{gyrotropicEc}
\rm E_c=\frac{\hbar c}{8\pi^3}\int_0^{\infty}d\xi \int_{-\infty}^{\infty} d^2 k_{\|} \,ln\, Det \left(I-R_+\tilde U_+R_-\tilde U_-\right),
\end{eqnarray}
where $\xi=-i\omega$ is the imaginary frequency.
This is our announced result for chiral Casimir energy with reflection matrix $\rm R_{\pm}$ and translation matrix $\rm \tilde U_{\pm}$ defined in chiral basis. 
\begin{table}
\caption{Phase velocities in Faraday materials and optically active materials}
\begin{tabular}{c|c|c|c}
\hline\hline
& & \multicolumn{2}{c}{phase velocity} \\
\cline{3-4}
\small{chirality}    & \small{propagating direction} & \small{Faraday materials} & \small{optically active materials} \\
\hline
\multirow{2}{*}{$\bold R$}      & $\bold +$   & \small{$\bar k_z+\delta k_z$} & \small{$\bar k_z+\delta k_z$}  \\

                     & $\bold -$    & \small{$\bar k_z-\delta k_z $}& \small{$\bar k_z+\delta k_z $}   \\\hline

 \multirow{2}{*}{$\bold L$} & $\bold +$& \small{$ \bar k_z-\delta k_z$} & \small{$\bar k_z-\delta k_z$}\\

 & $\bold -$& \small{$\bar k_z+\delta k_z $}&\small{$ \bar k_z-\delta k_z$}\\\hline
\end{tabular}
\end{table}

Note that the above derivation is based on optically active (P odd, T even) materials. In the following, we generalize the result to Faraday (P odd, T odd) materials. In Faraday materials,  where the velocity of photons depend on their propagating directions, the wave vectors in z direction of chiral photons have the following relation, i.e., $k_{zR}^{\pm}=k_{zL}^{\mp}=\bar k_z\pm\delta k_z$. Table I shows the phase velocity difference between Faraday materials and optically active materials. In this case, the translation matrices in two directions are not equivalent, i.e.,  $\rm \tilde U_+\neq \tilde U_-$.   Therefore, the above derivation does not hold at first sight. However, one can redefine chiral basis by interchange chiral basis in one direction. For example, one can defined a new set of chiral basis for photons propagating in the left direction, i.e., $\bold{\tilde e_{L-}}=\bold{e_{R-}}$ and $\bold{\tilde e_{R-}}=\bold{e_{L-}}$. In the new basis, we denote the new translation matrices in two directions with symbols  $\rm \tilde U_{+}^{\prime}$ and  $\rm \tilde U_{-}^{\prime}$. It is easy to show that $\rm \tilde U_{+}^{\prime} = \tilde U_{-}^{\prime}$, where $\rm \tilde U_{+}^{\prime}=\tilde U_{+}$ and $\rm \tilde U_{-}^{\prime}=I_A \tilde U_{-}I_A$. However, one should also note that the introduction of new basis will introduce two additional $\rm I_A$ matrices on two sides of translation operator $\rm I_A\tilde U_{-}^{\prime}I_A=\tilde U_{-}$. Therefore, the formula of chiral Casimir energy in Eqn. \eqref{gyrotropicEc} is general for Faraday materials.

Using our general formula, one can immediately obtain the classical results---the Casimir energy between two ideal, uncharged metal plates. In this case, the reflection matrices and translation matrices are $\rm r_{RR}=r_{LL}=0$, $\rm r_{LR}=r_{RL}=-1$, $\mathrm {U_{AB}=U_{BA}=}e^{ik_zl}\mathrm I$.  With these reflection matrices and translation matrices, one can obtain the Casimir force between two ideal metal plates, i.e., $\mathrm{E_0}=-\frac{\pi^2\hbar c}{720 l^3}$, where $l$ is the distance between the two plates.

\vspace{5mm}

\begin{center}
\textbf{${\rm\uppercase\expandafter{\romannumeral2}}$. Temperature dependence of chiral Casimir force.}
\end{center}

Usually, one can obtain the temperature dependence of Casimir energy via the substitution $\xi\mapsto \xi_n\equiv 2\pi n/\beta$ and $\frac{\hbar }{2\pi}\int d\xi \rightarrow \frac{1}{2\beta}\sum_{n=-\infty}^{\infty}$ in Eqn.\eqref{eq11}, where $\xi_n$ is the Matsubara frequency and $\rm \beta\equiv 1/k_B T$ \cite{bordag2}. Thus, the Casimir energy per area at finite temperature is given by
\begin{eqnarray}
\begin{split}
\mathrm{E_{c}(T)}=&\frac{1}{2\beta}\frac{1}{(2\pi)^2}\sum_{n=-\infty}^{\infty}\int_{-\infty}^{\infty} dk_x dk_y\,\, \ln\left[1+e^{-4\kappa l}-2e^{-2\kappa  l} \mathrm{cos}(2\mathcal V B l)\right]\\
=&\frac{1}{4\pi\beta}\sum_{n=-\infty}^{\infty}\int_0^{\infty} k_{\|} dk_{\|} \,\,\ln\left[1+e^{-4\kappa l}-2e^{-2\kappa l}\mathrm{cos}(2\mathcal V Bl)\right],
\end{split}
\end{eqnarray}
where $\kappa=\sqrt{\xi_n^2+k_{\|}^2}=\sqrt{\xi_n^2+k_x^2+k_y^2}$. In the classical limit, i.e., $\rm T\rightarrow \infty$, only the $n=0$ term domains. In this case, the Casimir energy is
\begin{eqnarray}\label{ect}
\mathrm{E_c(T)}=\frac{\mathrm{k_B T}}{4\pi} \frac{1}{l^2} \int_0^{\infty} x dx \,\, \ln \left[1+e^{-4x}-2e^{-2x} \mathrm{cos}(2\mathcal V(0) Bl)\right].
\end{eqnarray}
Notice that we assumed that the dielectric properties of materials A, B and C are not sensitive to the temperature. In the above calculation, we only kept the zeroth order. If we include the first order, i.e., sum over $\xi_0$ and $\xi_{\pm 1}$  terms, then, the temperature dependence of Casimir force is not linear any more.
One more comment on the classical limit $\rm T\rightarrow \infty$. This limit actually indicates that $\mathrm{k_B T}>>\hbar c k\sim \hbar c/l$. For instance, when two plates are separated about $\rm 10\,\mu m$, the high temperature limit corresponds to $\rm T>>100 K$. In the following, we numerically calculate the finite temperature Casimir force. We evaluate the summation up to $n=100$, i.e., $\sum_{n=-\infty}^{\infty}\approx \sum_{n=-100}^{100}$,  which is sufficiently accurate for temperature $T\geq 100 K$ at distance $l\geq 1\mu m$. Figure 6 (a) shows the chiral Casimir force at three different temperatures. As we can see from the figure, both enhanced and oscillating behavior are preserved at finite temperature.  Figure 6 (b) shows the chiral Casimir force at zero mass concentration $\rho=0$, i.e., without chiral material inserted  between two plates A \& B. The linearized behavior can be understood from Eqn. \eqref{ect}, as the denominator is $l^2$ instead of $l^3$. 
\begin{figure*}[!htb]
\centering
\includegraphics[height=4cm, width=11.5cm, angle=0]{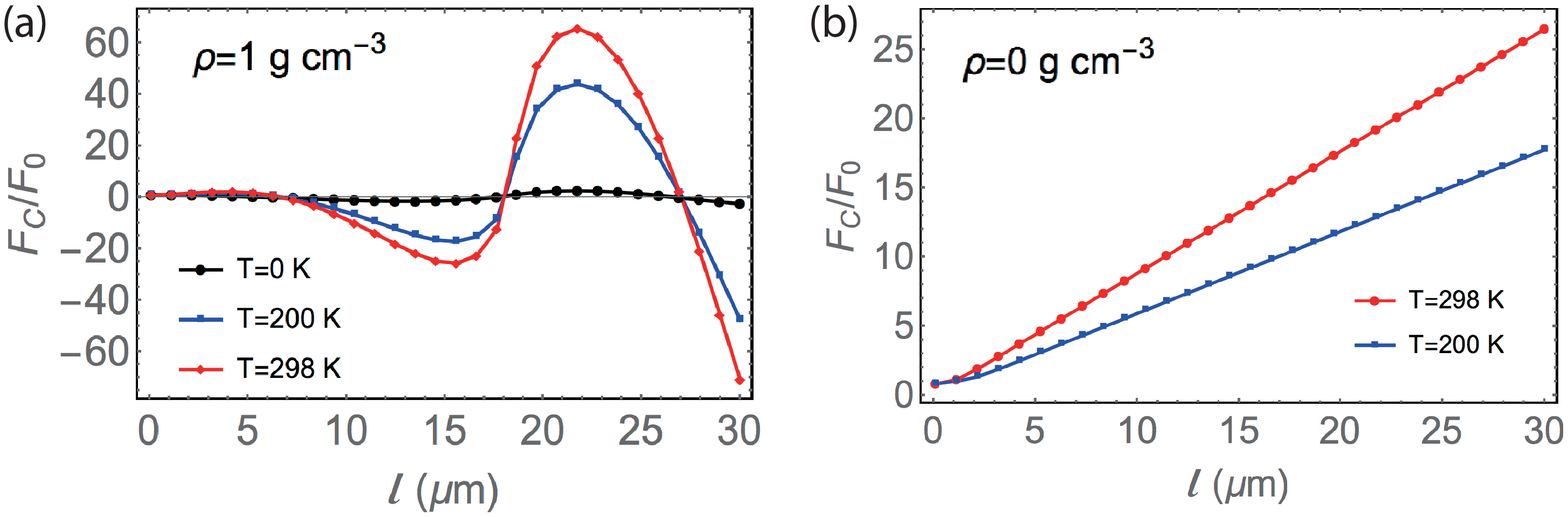}
\caption{Temperature dependence of chiral Casimir force.  The force is normalized with the classical Casimir force $F_0$, as is illustrated in the main text. (a) shows the chiral Casimir force at three different temperature $\rm T=0K$ (black curve),. $\rm T=200 K$ (blue curve),.$\rm T=298 K$ (red curve). (b) shows  the Casimir force at temperature $\rm T=298 K$ without chiral materials being inserted.  Parameters are set as: $\omega_p=10^{16} \rm s^{-1}$, $\gamma_0=2\times10^{6}\rm deg\,dm^{-1}g^{-1}cm^{3}$, $\gamma_1=0\,\rm deg\,dm^{-1}g^{-1}cm^{3}s^2$. The mass concentration is set as: $\rho=1 \,\rm g\, cm^{-3}$ in figure (a) and $\rho=0\, \rm g\, cm^{-3}$ in figure (b). \label{figure6}}
\end{figure*}

\begin{center}
\textbf{Supplementary References}
\end{center}

\end{widetext}


\begin{thebibliography}{}
\bibitem{Casimir}
H. Casimir,  On the attraction between two perfectly
conducting plates, Proc. K. Ned. Akad. Wet. \textbf{51}, 793 (1948).

\bibitem{Lifshitz}
E. M. Lifshitz, The theory of molecular attractive forces between solids,  Sov. Phys. JETP \textbf{2}, 73 (1956).

\bibitem{Dzyaloshinskii}
I. E. Dzyaloshinskii,  E. M. Lifshitz and L. P. Pitaevskii,  General theory of van der Waals' forces,  Adv. Phys. \textbf{10}, 165 (1961).

\bibitem{oken}
O. Kenneth and I. Klich,  Opposites Attract: A Theorem about the Casimir Force, Phys. Rev. Lett.  \textbf{97}, 160401 (2006).

\bibitem{Boyer}
T. H. Boyer, Van der Waals forces and zero-point energy for dielectric and permeable materials,  Phys. Rev. A 9, 2078 (1974);  O. Kenneth, \textit{et al}., Repulsive Casimir Forces, Phys. Rev. Lett. \textbf{89}, 033001 (2002).

\bibitem{Levin}
M. Levin,  A. P. McCauley,  A. W. Rodriguez,  M. T. Homer Reid,
and S. G. Johnson, Casimir repulsion between metallic objects in vacuum, Phys. Rev. Lett. \textbf{105}, 090403 (2010).

\bibitem{Leonhardt}
U. Leonhardt and T. G. Philbin, Quantum levitation by left-handed metamaterials, New J. Phys. \textbf{9}, 254
(2007).

\bibitem{jnmunday}
J. N. Munday, F. Capasso, and V. A. Parsegian, Measured long-range repulsive Casimir-Lifshitz forces, Nature (London) \textbf{457}, 170 (2009).

\bibitem{Tse2012}
W.-K. Tse and A. H. MacDonald, Quantized Casimir force,
Phys. Rev. Lett. \textbf{109}, 236806 (2012);
A. G. Grushin and A. Cortijo, Tunable Casimir Repulsion with Three-Dimensional Topological Insulators,
Phys. Rev. Lett. \textbf{106}, 020403 (2011);
P.  Rodriguez-Lopez and A. G. Grushin, Repulsive Casimir Effect with Chern Insulators, Phys. Rev. Lett. \textbf{112}, 056804 (2014);
J. H. Wilson, A. A. Allocca, and V. Galitski, Repulsive Casimir force between Weyl semimetals,
Phys. Rev. B \textbf{91}, 235115 (2015). 

\bibitem{rosa}
F. S. S. Rosa, D. A. R. Dalvit, and P. W. Milonni, Casimir-Lifshitz Theory and Metamaterials,
Phys. Rev. Lett. \textbf{100}, 183602 (2008);  I G Pirozhenko and A Lambrecht, Casimir repulsion and metamaterials, J. Phys. A: Math. Theor. \textbf{41}, 164015 (2008);  R. Zhao, J. Zhou, Th. Koschny, E. N. Economou, and C. M. Soukoulis, Repulsive Casimir Force in Chiral Metamaterials,
Phys. Rev. Lett.  \textbf{103}, 103602 (2009); V. Yannopapas and N. V. Vitanov,  First-Principles Study of Casimir Repulsion in Metamaterials, Phys. Rev. Lett. \textbf{103}, 120401 (2009).

\bibitem{supplement}
See the supplemental materials for details.

\bibitem{sma1}
M. T. Jaekel and S. Reynaud, Casimir force between partially transmitting mirrors, J. Phys. I (France) \textbf{1}, 1395
(1991);
A. Lambrecht, P. A. Maia Neto, and S. Reynaud, The Casimir effect within scattering theory, New J.
Phys. \textbf{8}, 243 (2006).

\bibitem{sma2}
T. Emig, A. Hanke, R. Golestanian, and M. Kardar, Probing the Strong Boundary Shape Dependence of the Casimir Force, Phys. Rev. Lett. \textbf{87}, 260402 (2001);
S. J.  Rahi, T. Emig, N. Graham, R. L. Jaffe, and M. Kardar, Scattering theory approach to electrodynamic Casimir forces,
Phys. Rev. D \textbf{80}, 085021 (2009).

\bibitem{sma3}
C.  Genet, A. Lambrecht, and S. Reynaud, Casimir force and the quantum theory of lossy optical cavities, Phys. Rev. A \textbf{67}, 043811 (2003);
S. Y.  Buhmann, D. T. Butcher and S. Scheel, Macroscopic quantum electrodynamics in nonlocal and nonreciprocal media, New J. Phys. \textbf{14}, 083034 (2012).

\bibitem{bordag1}
M. Bordag, U. Mohideen, and V. M. Mostepanenko, 
New Developments in the Casimir Effect, Phys.
Rep. \textbf{353}, 1 (2001).

\bibitem{footnote2}
Reflecting lack of time reversal symmetry, photons with the same chirality have different phase velocity when propagating in different directions. More discussion on symmetry and phase velocity in gyrotropic materials can be found in the supplemental materials \cite{supplement}.

\bibitem{footnote3}
The reason is the left (right)-hand polarized electromagnetic waves will become right (left)-hand polarized.
Because, for perfect conductor $\rm E_x\rightarrow -E_x$ and $\rm E_y\rightarrow -E_y$ with $k_z\rightarrow -k_z$. Therefore, the chirality of photons changes after being reflected.

\bibitem{crassee}
I. Crassee, \textit{et al}., Giant Faraday rotation in single-and multilayer graphene, 
Nat. Phys. \textbf{7}, 48 (2011);
M. C. Sekhar, M. R. Singh, S. Basu, and S. Pinnepalli,  Giant Faraday rotation in $\rm Bi_xCe_{3-x}Fe_5O_{12}$ epitaxial garnet films, Opt. Express, \textbf{20}, 9624 (2012);
S. Vandendriessche, \textit{et al}., Giant faraday rotation in mesogenic organic molecules, Chem. Mater. \textbf{25},  1139 (2013).

\bibitem{plum}
E. Plum and N. I. Zheludev, Chiral mirrors,  Appl. Phys. Lett. \textbf{106}, 221901 (2015);
V.  A. Fedotov, S. L. Prosvirnin, A. V. Rogacheva, and N. I.
Zheludev,  Mirror that does not change the phase of reflected waves, Appl. Phys. Lett. \textbf{88}, 091119 (2006).

\bibitem{vandendriessche2}
S. Vandendriessche, V. K. Valev, and T. Verbiest, Faraday rotation and its dispersion in the visible region for saturated organic liquids, Phys. Chem. Chem. Phys.  \textbf{14}, 1860 (2012).

\bibitem{emile}
M.  Kuwata-Gonokami, \textit{et al}.,  Giant optical activity in quasi-two-dimensional planar nanostructures, Phys. Rev. Lett. \textbf{95}, 227401 (2005);
J.  Emile, \textit{et al}., Giant optical activity of sugar in thin soap films, J Colloid Interface Sci.  \textbf{408}, 113 (2013);
S. Takahashi, \textit{et al}., Giant optical rotation in a three-dimensional semiconductor chiral photonic crystal,
Opt. Express \textbf{21}, 29905 (2013).


\end{thebibliography}

\begin{thebibliography}{}

\bibitem{raabe}
C. Raabe and D.-G. Welsch, Casimir force acting on magnetodielectric bodies embedded in media,
Phys. Rev. A \textbf{71}, 013814 (2005).

\bibitem{buhmann}
S. Y. Buhmann, D. T. Butcher, and S. Scheel, Macroscopic quantum electrodynamics in nonlocal and nonreciprocal media, New. J. Phys. \textbf{14}, 083034 (2012).

\bibitem{sfuchs}
S. Fuchs,  \textit{et al.}, Casimir-Lifshitz force for nonreciprocal media and applications to photonic topological insulators, Phys. Rev. A \textbf{96}, 062505 (2017).

\bibitem{bordag2}
M. Bordag, U. Mohideen, and V. M. Mostepanenko, 
New Developments in the Casimir Effect, Phys.
Rep. \textbf{353}, 1 (2001).

\end{thebibliography}
\end{document}